\documentclass[10pt, paper=a4, UKenglish]{article}
\usepackage{graphicx}
\usepackage{lineno}
\def\Title#1{\begin{center} {\Large #1 } \end{center}}
\def\Author#1{\begin{center}{ \sc #1} \end{center}}
\def\Address#1{\begin{center}{ \it #1} \end{center}}

\newcommand\pubblock{\rightline{\begin{tabular}{l} Proceedings of the CTD/WIT 2019\\ \pubnumber\\
         \pubdate  \end{tabular}}}

\newenvironment{Abstract}{\begin{quotation} \begin{center} 
             \large ABSTRACT \end{center}\bigskip 
      \begin{center}\begin{large}}{\end{large}\end{center} \end{quotation}}

\newenvironment{Presented}{\begin{quotation} \begin{center} 
             PRESENTED AT\end{center}\bigskip 
      \begin{center}\begin{large}}{\end{large}\end{center} \end{quotation}}

\def\beq{\begin{equation}}
\def\eeq#1{\label{#1}\end{equation}}
\def\eeqn{\end{equation}}

\def\beqa{\begin{eqnarray}}
\def\eeqa#1{\label{#1}\end{eqnarray}}
\def\eeqan{\end{eqnarray}}

\let\bar=\overbar

\def\Dslash{\not{\hbox{\kern-4pt $D$}}}
\def\dslash{\not{\hbox{\kern-2pt $\del$}}}

\def\msb{{\bar{\ssstyle M \kern -1pt S}}}

\usepackage{color}
\usepackage{lineno}
\usepackage{subfig}
\usepackage{hyperref}

\newcommand\pubnumber{PROC-CTD19-036}

\newcommand\pubdate{\today}

\def\affiliation{
On behalf of the ATLAS collaboration
}

\newcommand{\conference}{Connecting the Dots and Workshop on Intelligent Trackers (CTD/WIT 2019)\\
Instituto de F\'isica Corpuscular (IFIC), Valencia, Spain\\ 
April 2-5, 2019 \newline \newline
Copyright 2019 CERN for the benefit of the ATLAS Collaboration. CC-BY-4.0 license}

\usepackage{fancyhdr}
\definecolor{mygrey}{RGB}{105,105,105}
\begin{document}

% uncomment the following line for adding line numbers
% \linenumbers

% large size for the first page
\large
\begin{titlepage}
\pubblock

%% Change the title, name, abstract
%% Title 
\vfill
\Title{First tracking performance results from the ATLAS Fast TracKer}
\vfill

%  if you need to add the support use this, fill the \support definition above. 
%  \Author{FIRSTNAME LASTNAME \support}
\Author{Benjamin Hooberman}
\Address{\affiliation}
\vfill

\begin{Abstract}
Particle physicists at the Large Hadron Collider investigate the properties of matter at subatomic length scales
by colliding together bunches of high-energy protons and observing the decay products of 
the collisions. ATLAS is one of two general-purpose detectors that reconstruct the interactions and, as part of a wide range of 
physics goals, measure the production of Higgs bosons and searches for exotic new phenomena including supersymmetry, extra dimensions
of spacetime, and dark matter. \newline

Selecting the interesting collision events using hardware- and software-based triggers is a major challenge that
will become more difficult as the luminosity increases in future data. The ATLAS Fast TracKer (FTK) 
is a custom electronics system that performs fast hardware-based tracking of charged particles for use in trigger decisions. 
In 2018, two FTK ``Slices'' covering portions of the ATLAS detector were installed and commissioned using proton-proton collisions,
to prepare for physics data-taking in Run 3. The
FTK track-finding and track-fitting strategies and the tracking performance for the FTK Slices are presented. 
Strategies for coping with changing beamspot and other conditions in future data are discussed. 
A strategy for triggering on displaced tracks from long-lived particles is also presented.
\end{Abstract}

\vfill

% DO NOT CHANGE!!!
\begin{Presented}
\conference
\end{Presented}
\vfill
\end{titlepage}
\def\thefootnote{\fnsymbol{footnote}}
\setcounter{footnote}{0}
%

% normal size for the rest
\normalsize 

%% Your paper should be entered below. 

\section{Motivation}
\label{intro}

% feel free to set the height or width as you prefer
\begin{figure}[!t]
  \centering
  \includegraphics[width=1.0\linewidth]{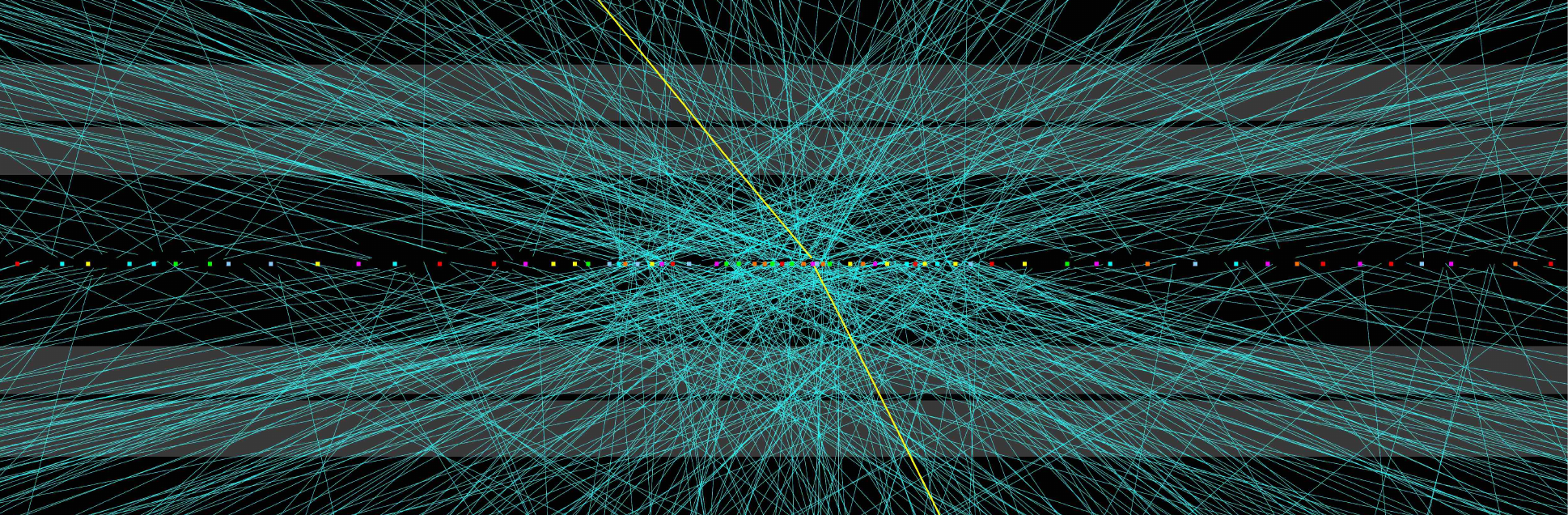} 
  \caption{\label{fig:display} Event display of the interaction region in ATLAS proton-proton collisions
    with $\langle \mu \rangle=60$ average pile-up interactions~\cite{cite:display}. Green lines
    are charged particle tracks and colored dots are interaction vertices. 
    Two high momentum tracks from the primary vertex are indicated by yellow lines.
    %Right: the dependency of tracking reconstruction wall time per event on the average 
    %number of interactions per bunch crossing ($<\mu>$)~\cite{}. 
}  
\end{figure}

The Large Hadron Collider (LHC) probes physics on subatomic scales by colliding together bunches of 100 billion protons 
40 million times per second. As shown in Figure~\ref{fig:display}, the flux of particles produced in the collisions is large
due to multiple overlapping pile-up interactions that are superimposed on the collision of interest. In LHC Run 3,
which will take place from 2021-2023, an average pile-up ($\langle \mu \rangle$) of 60 interactions per bunch crossing are expected. 
By reconstructing the charged particle tracks and using them to identify the locations of the primary interaction vertex 
and pile-up vertices, the particles produced in pile-up interactions can be identified and suppressed during reconstruction.
Software-based tracking is too slow to perform full-event tracking at the full Level-1 accept rate of 100 kHz within
the latency constraint of order 100 $\mu$s. A hardware-based solution is therefore needed to perform full-event tracking
for use in trigger decisions. The ATLAS Fast TracKer (FTK)~\cite{cite:FTK} is a custom electronics system that uses custom 
Associative Memory (AM) chips and Field-Programmable Gate Arrays (FPGA)
to perform fast tracking in hardware, for use in trigger decisions. The system is currently being installed 
and commissioned in ATLAS and will be used to enhance the physics potential of the LHC Run 3. 

\section{ATLAS Trigger and Data Acquisition}
\label{section}

\begin{figure}[!t]
  \centering
  \subfloat[]{\includegraphics[width=0.45\linewidth]{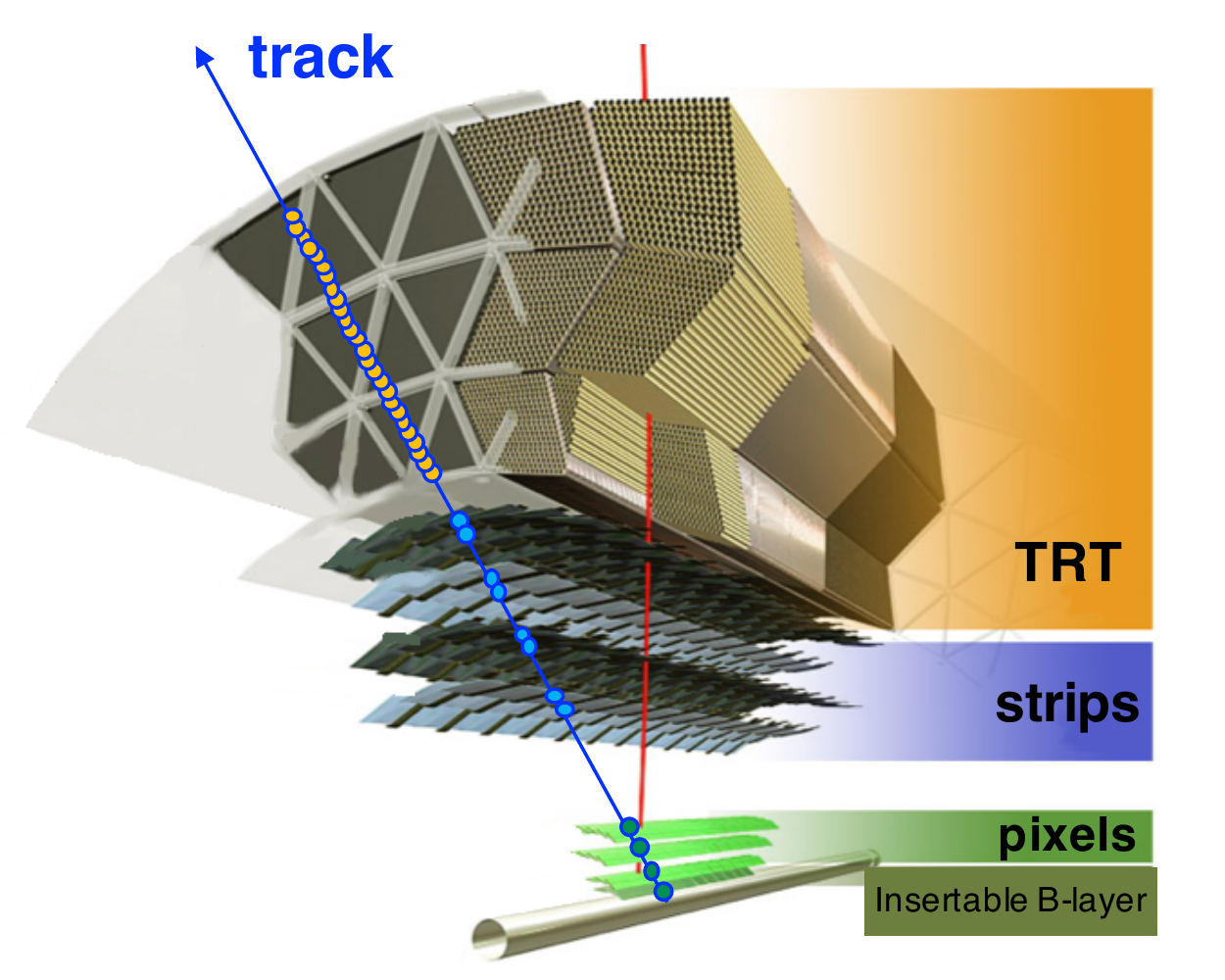}}
  \qquad
  \subfloat[]{\includegraphics[width=0.45\linewidth]{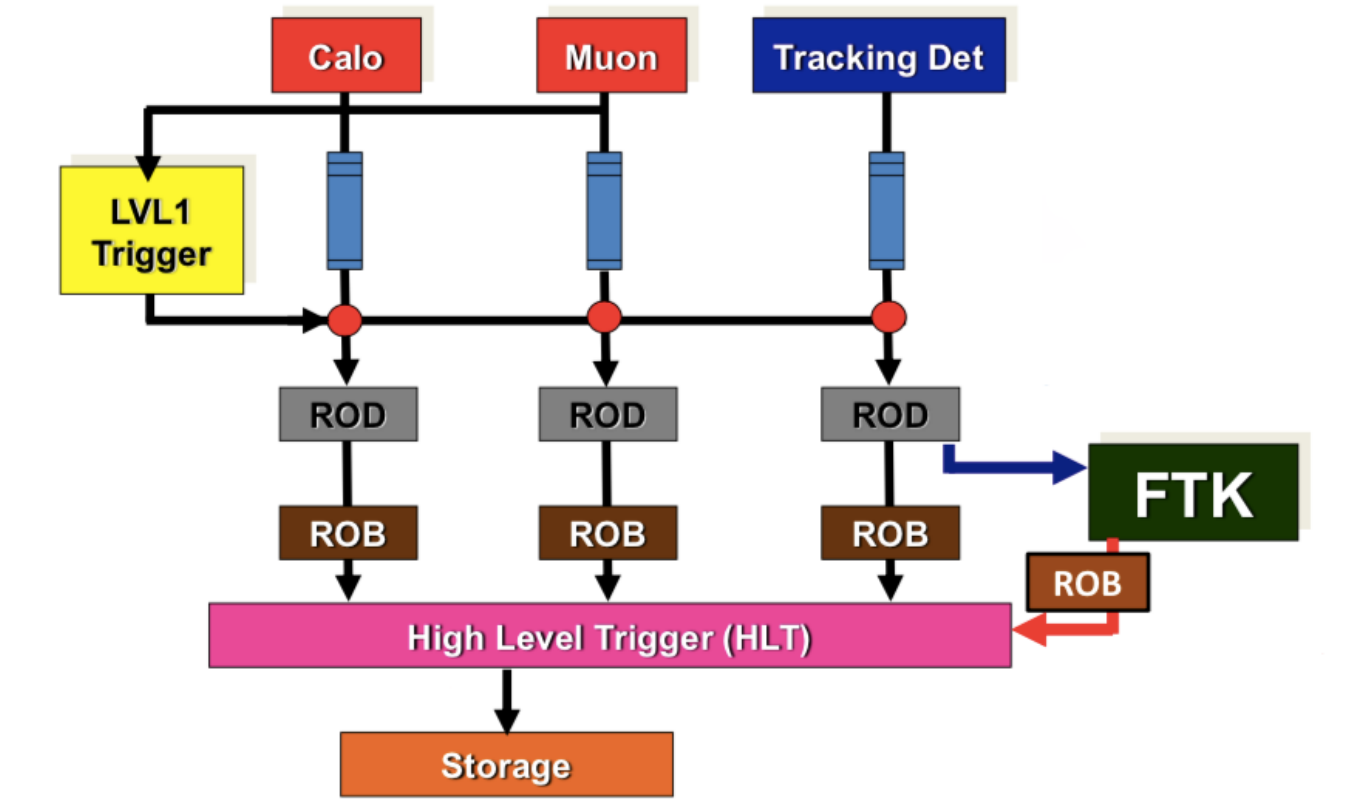}}
  \caption{(a) Section of the ATLAS inner tracking detector.
    The transition radiation tracker (TRT), strips, pixels, and Insertable B-layer (IBL) are indicated. 
    (b) Diagram of the ATLAS trigger and data acquisition system including the FTK.}
  \label{fig:pictures}
\end{figure}

The ATLAS detector~\cite{cite:ATLAS} is a multipurpose particle physics detector with nearly 4$\pi$ coverage in solid angle 
around the collision point\footnote{ATLAS uses a right-handed coordinate system with its origin at the nominal interaction 
point (IP) in the centre of the detector 
and the $z$-axis along the beam pipe. The $x$-axis points from the IP to the centre of the LHC ring, and the $y$-axis points upwards.
Cylindrical coordinates $(r, \phi)$ are used in the transverse plane, $\phi$ being the azimuthal angle around the $z$-axis. The 
pseudorapidity is defined in terms of the polar angle $\theta$ as $\eta = −- \ln \tan(\theta/2)$. Angular distance is measured in 
units of $\Delta R \equiv \sqrt{ (\Delta\eta)^2 + (\Delta\phi)^2 }$.
The transverse momentum, $p_\mathrm{T}$, is defined with respect to the beam axis ($x–y$ plane).}.
It consists of an inner tracking detector (ID), surrounded by a superconducting
solenoid providing a 2T axial magnetic field, a system of calorimeters, and a muon spectrometer (MS) incorporating three 
large superconducting toroid magnets.
The ID provides charged-particle tracking in the range $|\eta| < 2.5$. 
It consists of approximately 100 million double-sided strips providing stereo and axial 1-dimensional hits
and pixels providing 2-dimensional hits, as well as a transition radiation tracker (TRT), as shown in  Figure~\ref{fig:pictures}(a).
During the LHC shutdown between Run 1 (2010--2012) 
and Run 2 (2015--2018), a new innermost layer of silicon pixels was added, which improves the track impact 
parameter resolution, vertex position resolution and b-tagging performance~\cite{cite:IBL}.

The ATLAS trigger and data acquisition (TDAQ) system~\cite{cite:trigger} consists of two tiers: 
a hardware-based Level-1 trigger selects approximately 100 kHz
of interesting events from the LHC bunch crossing rate of 40~MHz, and a software-based 
high-level trigger (HLT) runs reconstruction and calibration software similar to the offline reconstruction for events accepted by the
Level-1 trigger, reducing the event rate to about 1 kHz on average for offline storage and analysis. 
No tracking is possible in the Level-1 trigger, which uses data from the calorimeters and MS but not the tracking detectors.
Tracking in limited regions-of-interest (ROI) is possible in the HLT, but full-event tracking is not possible due to 
limited HLT computing resources. The FTK reads in tracker data for events passing the Level-1 accept and 
reconstructs tracks with $p_\mathrm{T}>1$ GeV over the full detector. These tracks are provided at the start of the HLT step,
alleviating the burden on HLT computing resources. The tracks can either be used directly or else quickly refit in the HLT.

\section{The ATLAS Fast TracKer}
\label{sec:tracking}

The FTK uses custom AM chips and FPGAs to quickly 
reconstruct the trajectories of charged particles.
The FTK system is highly parallel, with tracking performed simultaneously
in 64 independent $\eta-\phi$ ``towers''. Track-finding is performed by first grouping pixel and strip hits into
coarse resolution ``superstrips'', and then comparing these superstrips to a pre-computed set of about 1 billion
track ``patterns'' representing collections of hits that are consistent with the path of a charged particle through 
the ATLAS tracking detector layers.
Finally, the track parameters are estimated from the hit positions in the tracking detector layers using
a linearized approximation given by $p_i = \Sigma_j C_{ij} x_j + q_i$. Here $p_i$ are the track
parameters ($p_\mathrm{T}$, $\eta$, $\phi$, longitudinal impact parameter $d_0$, transverse impact parameter $z_0$,
and track goodness-of-fit $\chi^2$), $x_j$ are the relative hit positions in each layer, and $C_{ij}$ and $q_i$ are constants
that are defined for each FTK ``sector'', defined by a set of $\sim1$~cm$^2$ silicon sensor modules, one in each layer.
The patterns as well as the sectors and constants are determined from samples of fully-simulated muons.

The FTK hardware consists of five main electronics boards. Strip and pixel data are fed to the Input Mezzanine (IM),
which clusters the hits and sends them to the Data Formatter (DF).
The DF formats the clustered hits and sends them to the auxiliary board (AUX). The AUX performs the first tracking stage using 
8 of the available 12 tracking detector layers, by comparing the hits to patterns stored in custom AM chips.
These 8-layer track candidates are sent to the Second Stage Board (SSB) along with DF data from the four remaining
layers. The SSB extrapolates the 8-layer tracks to the four additional layers and adds hits to the tracks, estimates
the track parameters, and removes duplicate tracks. The 12-layer tracks are then sent to the FTK-to-Level-2 Interface Crate (FLIC),
where they are formatted and sent to the detector Readout System (ROS). 

\section{Data and simulation}
\label{sec:datasim}

% feel free to set the height or width as you prefer
\begin{figure}[!t]
  \centering
  \includegraphics[width=0.6\linewidth]{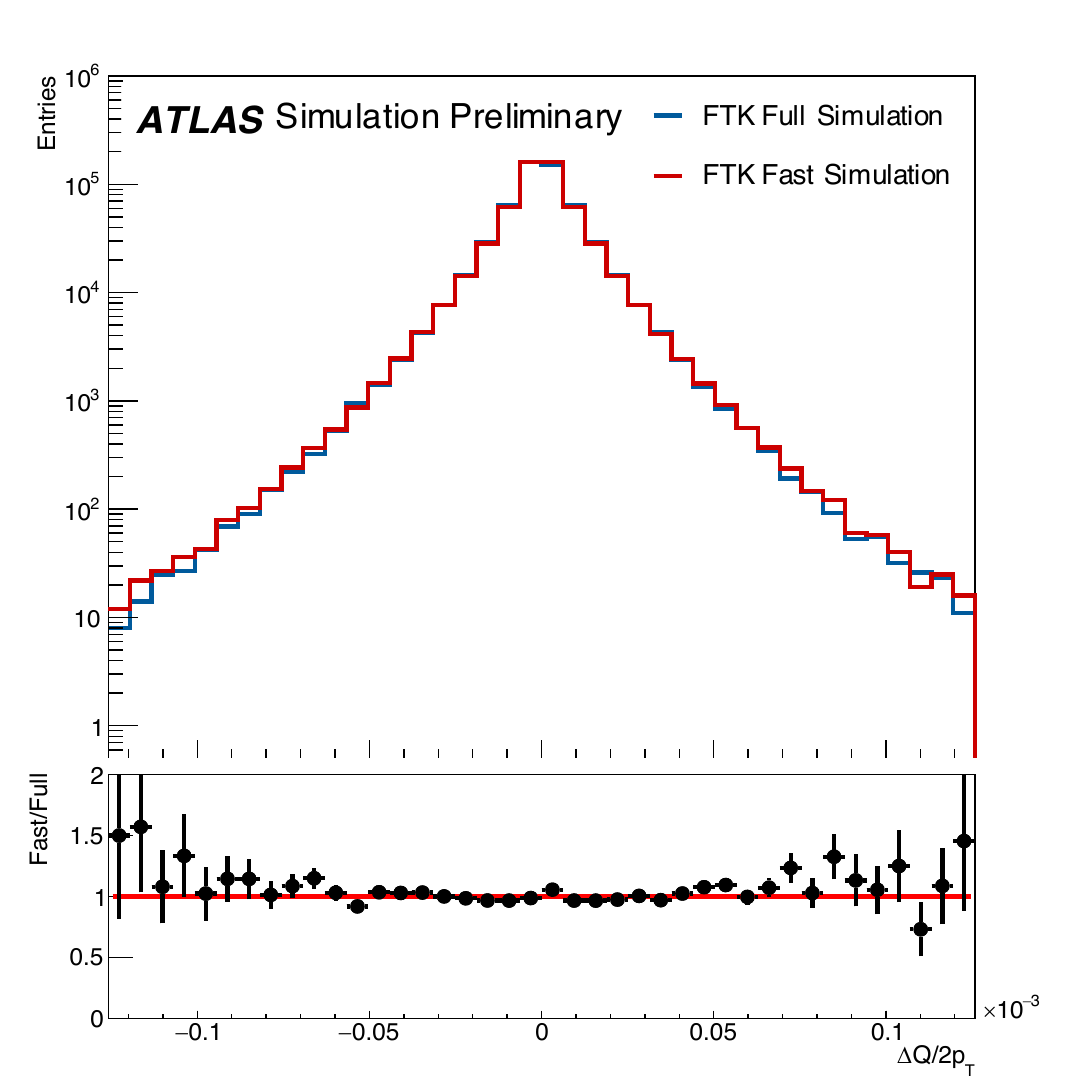} 
  \caption{\label{fig:fastsim}  
    The resolution of the FTK track curvature, from functional emulation of FTK using FTKSim (FTK Full Simulation) and a 
    parameterized uncertainty model (FTK Fast Simulation)~\cite{cite:twiki}. The difference between the FTK reconstructed and true values of 
    track charge divided by twice the track transverse momentum ($\Delta Q/2p_\mathrm{T}$) is plotted. A fully simulated sample of 
    muons with longitudinal impact parameter $|z_0| < 110$ cm and pseudo-rapidity $|\eta| < 2.5$ is used. 
    The resolution for the fast simulation is modeled as a double gaussian derived from full simulation. 
}  
\end{figure}

In 2018, 2 of the 64 FTK towers were instrumented with FTK ``Slices'' and proton-proton
commissioning data were collected.
The FTK Slice in Tower 40 ($0<\eta<1.5, 2.4 < \phi < 2.8$) implemented the first tracking stage to produce
8-layer track candidates, while the Slice in Tower 22 ($-1.5 < \eta < 0, 1.6 < \phi < 2.0$)
implemented full 12-layer track processing using a complete set of FTK boards.
The Tower 22 slice ran stably for two hours in a special high pile-up run ($\langle \mu \rangle=82$) in October 2018
and wrote out tracks to a special ATLAS data stream for trigger development and rate predictions.
%A single DF with partial coverage of Tower 22 was used and approximately 0.5M tracks were collected.
%FTK tracks with Insertable B-layer (IBL) hits were excluded, due to a FTK module ordering problem that
%caused incorrect hit positions in the run. The cause is understood and the fix is being implemented.

Two types of simulation are used to model the FTK track processing. A full functional emulation called FTKSim 
has been implemented in C++ code. Full simulation is used to train the sectors and constants and patterns
and to validate the firmware using bit-level comparisons to FTK tracks. This step is too slow (approximately 600 HS06 seconds per event)
to be useful for large-scale Monte Carlo sample productions. A fast simulation has therefore been developed
that uses efficiency weights and track parameter smearing to quickly emulate the FTK tracking performance. 
Track parameter resolutions are extracted from samples of muons simulated with FTKSim and separated into 
bins of track $p_\mathrm{T}$, $\eta$, and number
of IBL hits. For each bin, double Gaussian resolution functions are fit to the distribution of 
each of the five track parameter residuals, i.e., the reconstructed minus true values for each parameter.
The fast simulation is validated by applying the smearing functions to a simulated muon sample and comparing the resulting 
residuals to those obtained from full simulation. An example showing the modeling of the track curvature is presented
in Figure~\ref{fig:fastsim}. The fast simulation provides good modeling of the residuals in both the core and tails of the
resolution function.

\section{Tracking performance}
\label{sec:performance}

The tracking performance results from the FTK Tower 22 Slice are presented in this section.
For all results, a single Data Formatter (DF) board 
provides partial coverage of Tower 22, which spans $-1.5 < \eta < 0$ and $1.6 < \phi < 2.0$. The data cover a subset of 
380,000 events from Run 364485, a special high pile-up run with average interactions per crossing $\langle \mu \rangle = 82$, collected in 
October 2018. FTK tracks with Insertable B-layer (IBL) hits were excluded, due to a FTK module ordering problem that caused 
incorrect hit positions in the run. The cause is understood and the fix is being implemented. FTK tracks are matched to 
offline tracks within $\Delta R < 0.02$.

\begin{figure}[!t]
  \centering
  \subfloat[]{\includegraphics[width=0.3\linewidth]{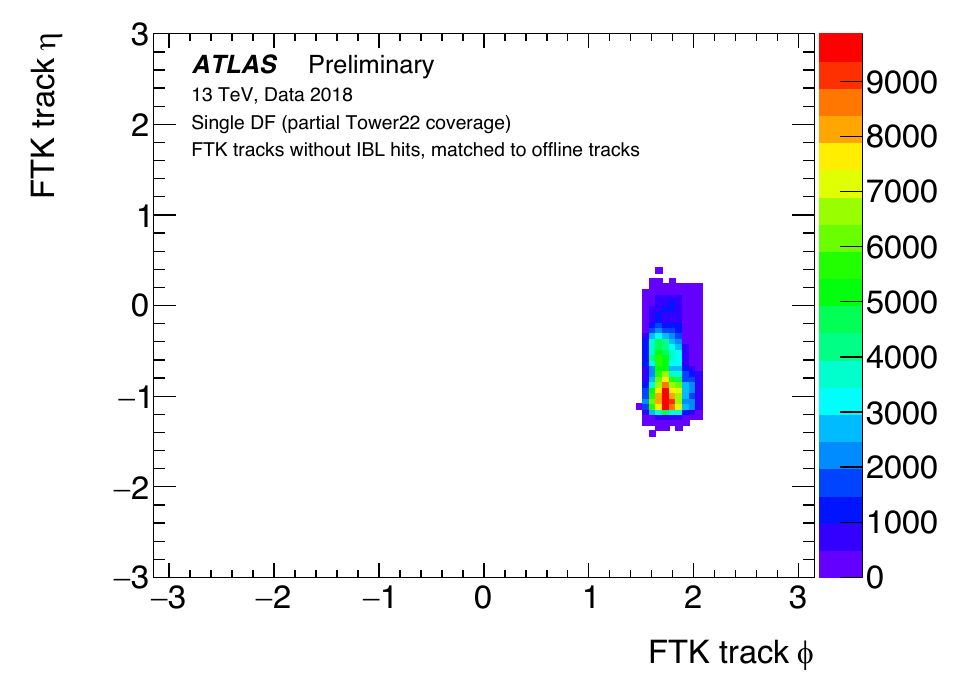}}
  \qquad
  \subfloat[]{\includegraphics[width=0.3\linewidth]{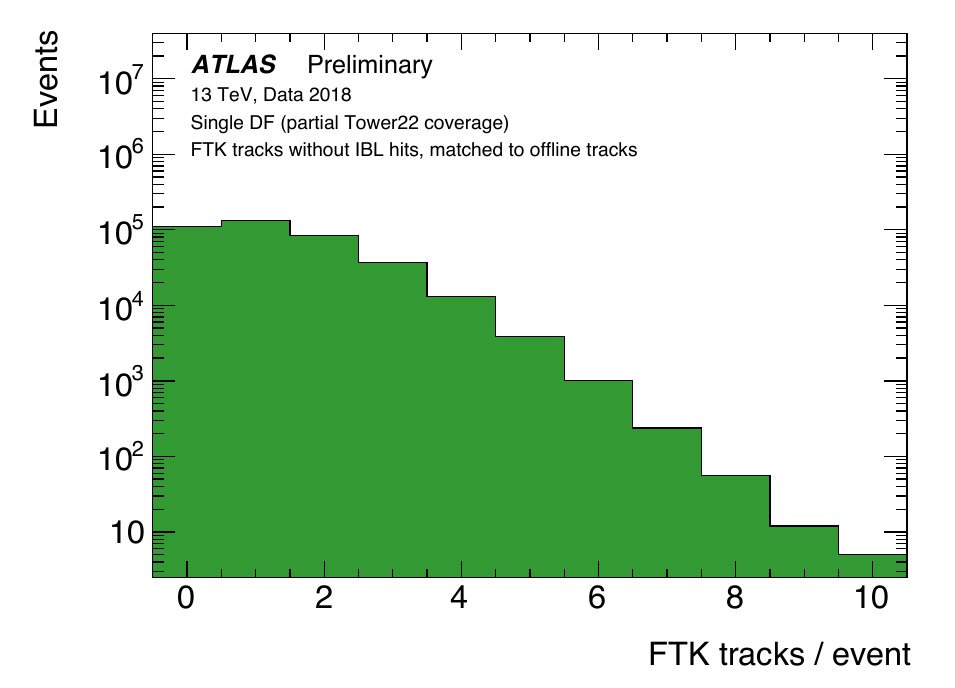}}
  \qquad
  \subfloat[]{\includegraphics[width=0.3\linewidth]{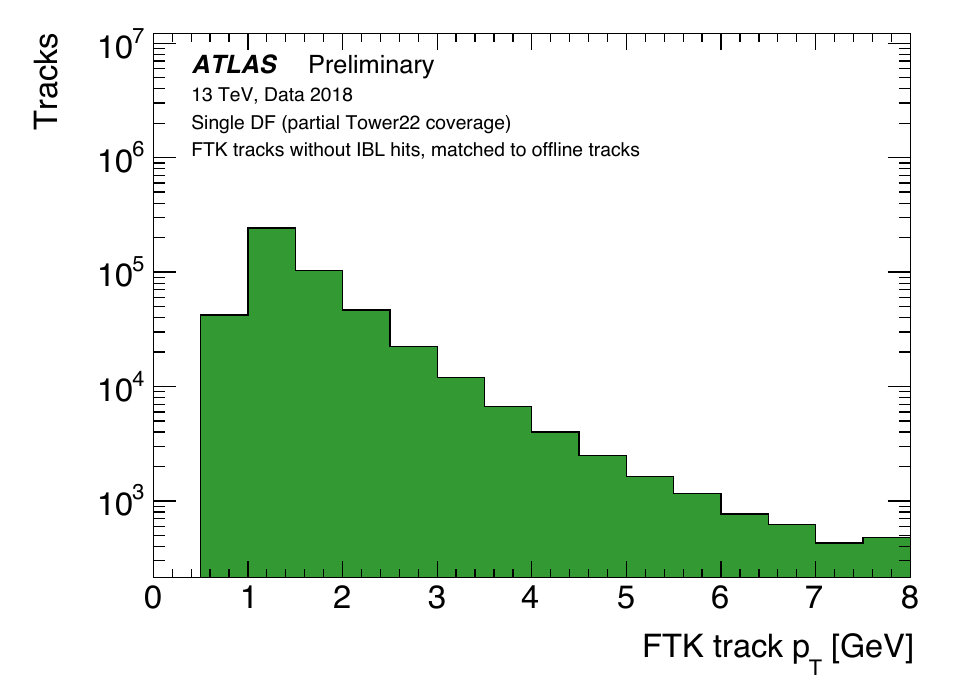}}
  \caption{
    The distribution of FTK (a) track pseudo-rapidity ($\eta$) vs. azimuthal angle ($\phi$), (b) number of tracks
    produced per event, and (c) track transverse momentum ($p_\mathrm{T}$), from the FTK Tower 22 Slice~\cite{cite:twiki}. } 
  \label{fig:tracks}
\end{figure}

% feel free to set the height or width as you prefer
\begin{figure}[!b]
  \centering
  \includegraphics[width=0.6\linewidth]{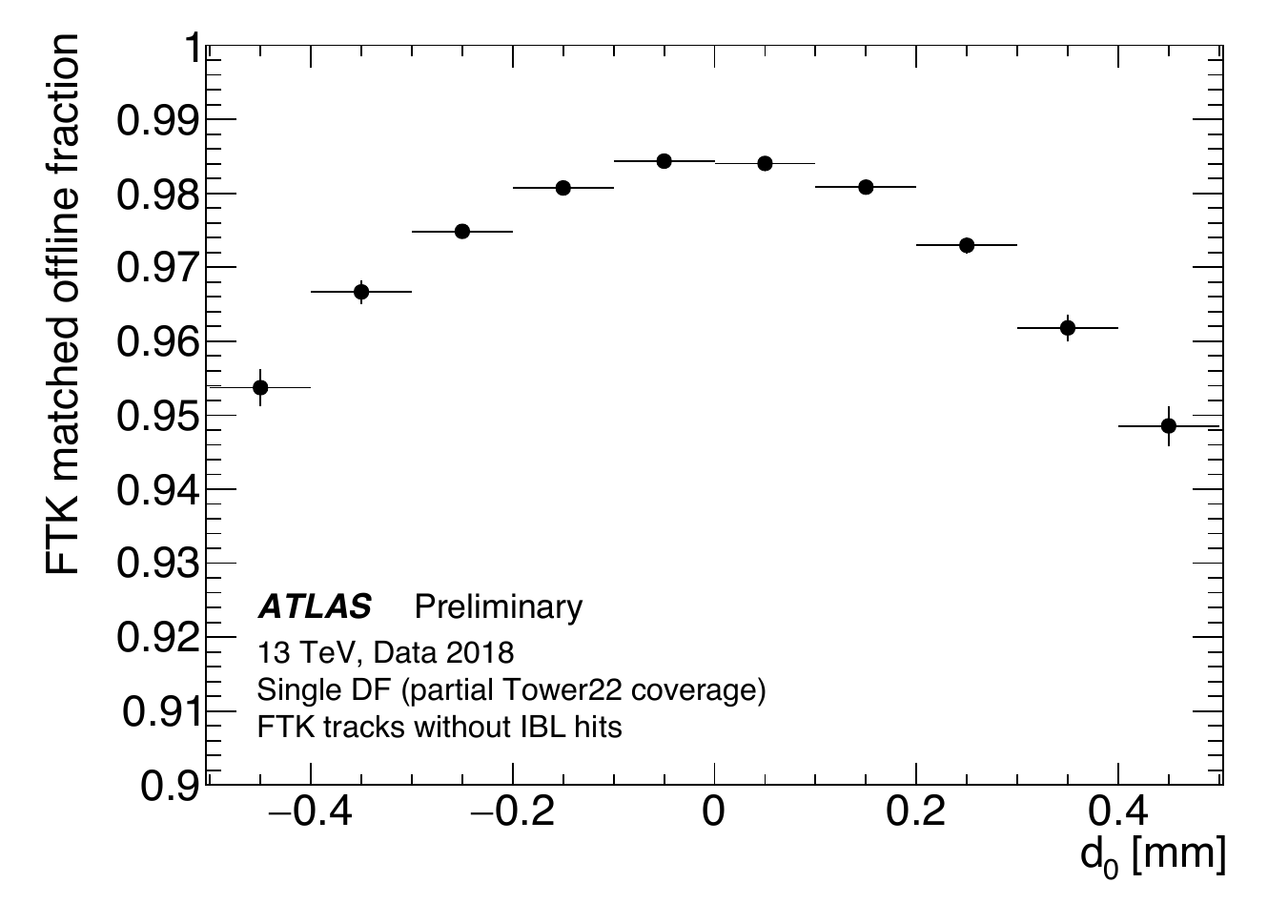} 
  \caption{\label{fig:purity} The fraction of FTK tracks matched to selected offline tracks, as a function of the 
    FTK track transverse impact parameter $d_0$, from the FTK Tower 22 Slice~\cite{cite:twiki}. 
}  
\end{figure}

\begin{figure}[!t]
  \centering
  \subfloat[]{\includegraphics[width=0.3\linewidth]{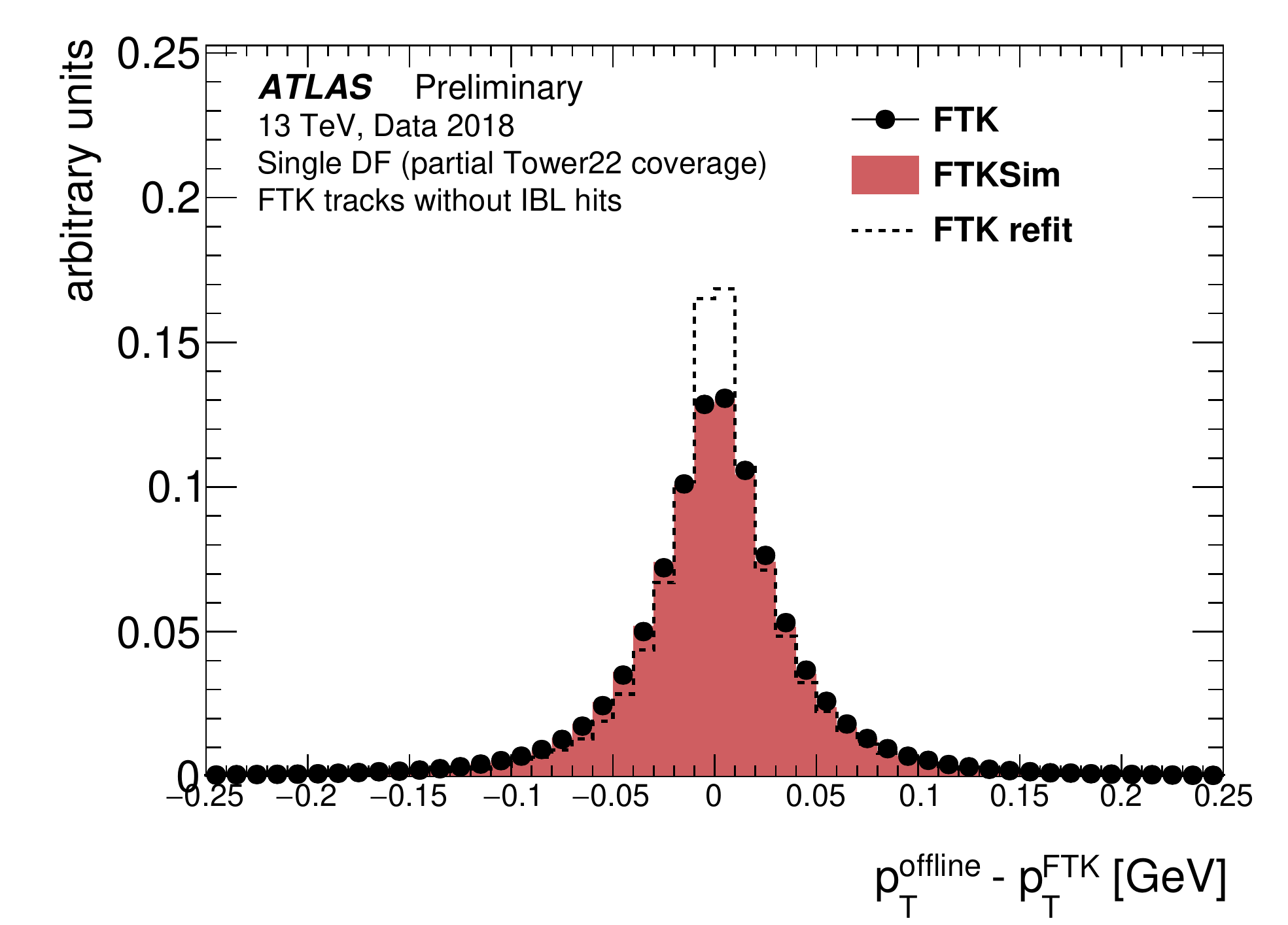}}
  \qquad
  \subfloat[]{\includegraphics[width=0.3\linewidth]{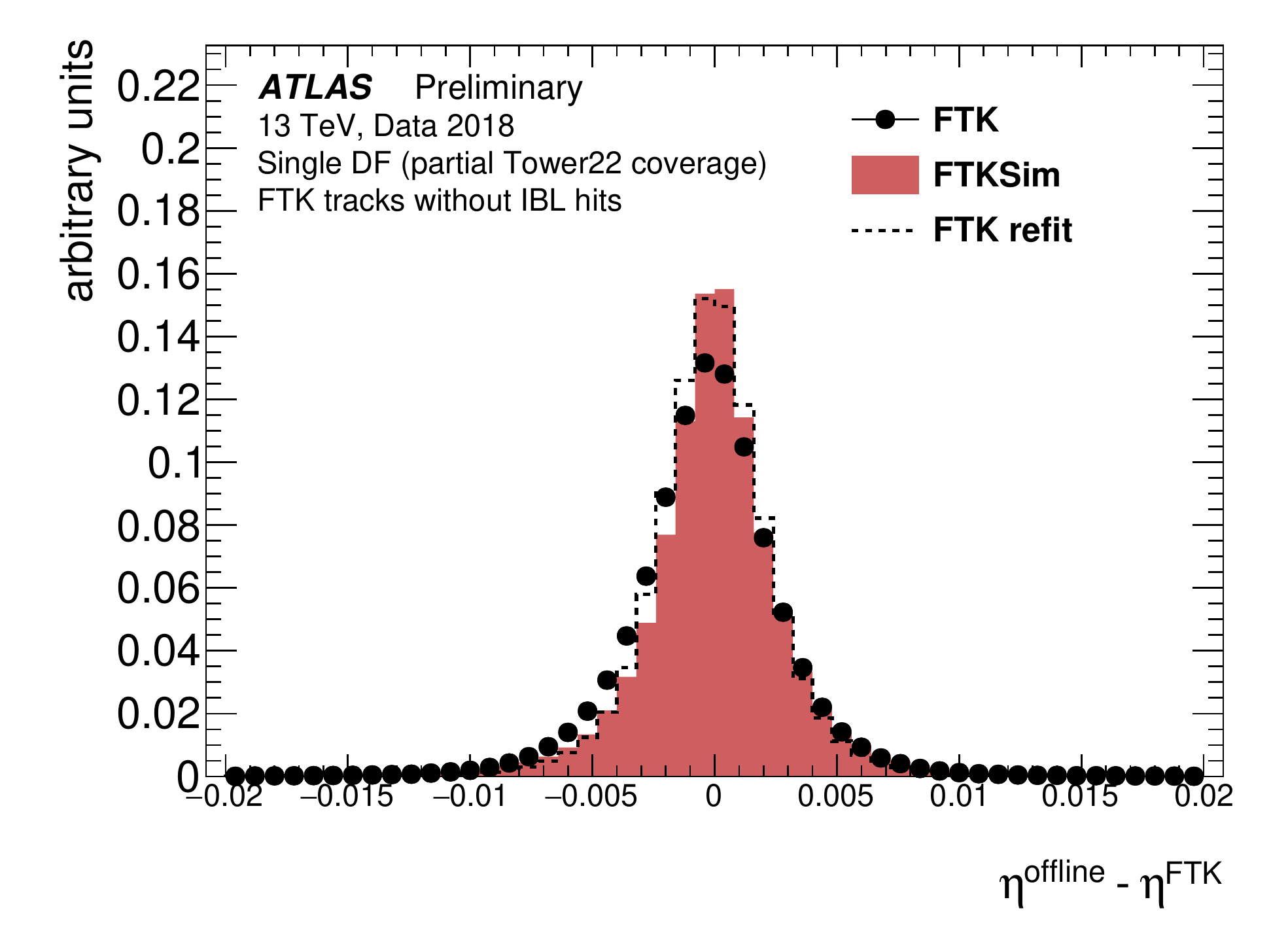}}
  \qquad
  \subfloat[]{\includegraphics[width=0.3\linewidth]{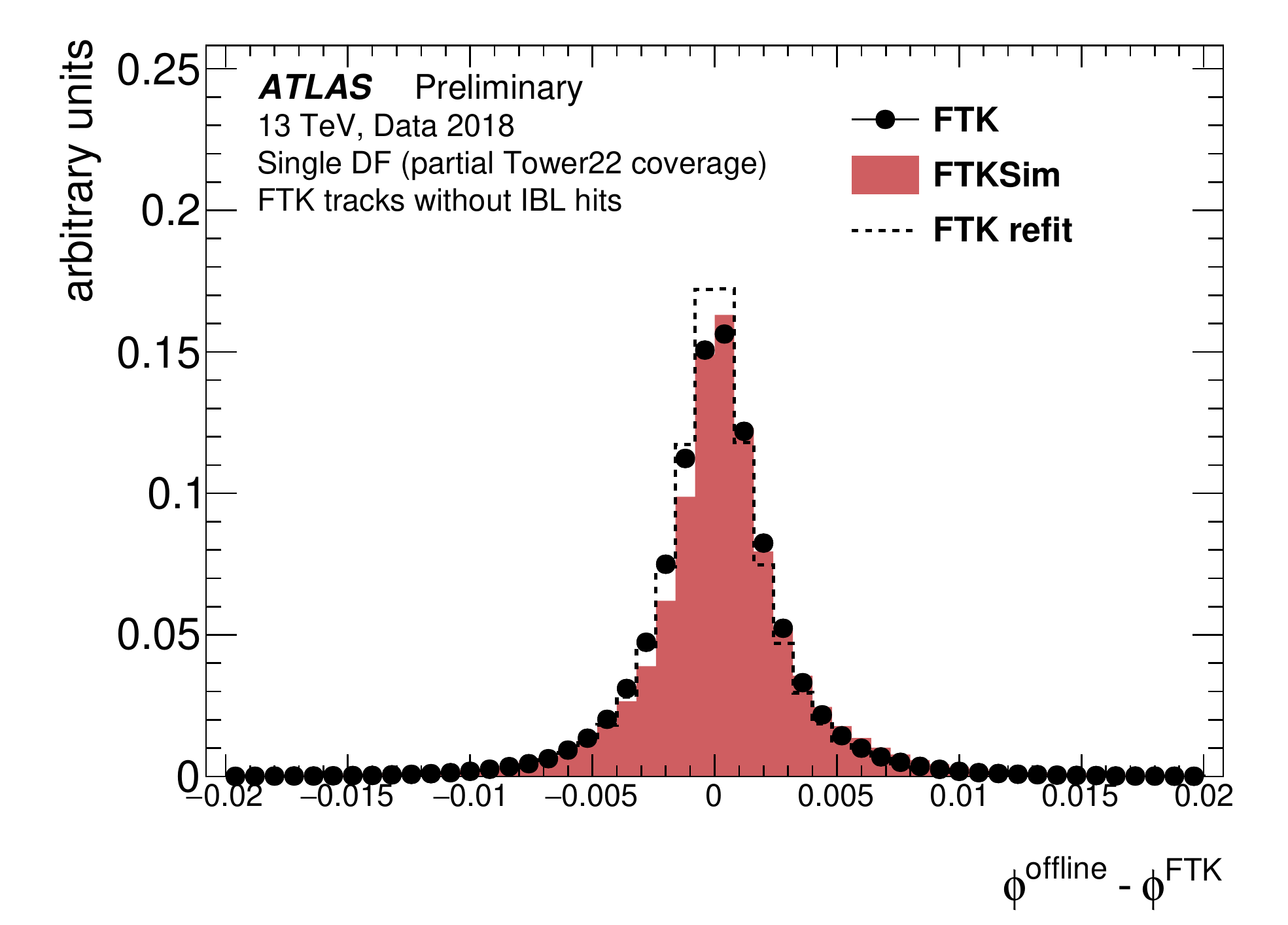}}
  \caption{
    FTK vs. offline track residuals from the FTK Tower 22 Slice for (a) $p_\mathrm{T}$, (b) $\eta$, and (c) $\phi$~\cite{cite:twiki}. The difference
    between the FTK and matched offline track pT are shown for (black dots) the FTK slice, (shaded red histogram) 
    functional emulation of the FTK slice using FTKSim, and (dashed line) FTK refit. } 
  \label{fig:resolutions}
\end{figure}

Figure~\ref{fig:tracks} presents distributions of the 0.5M 12-layer tracks collected in the Tower 22 Slice.
The distribution of track $\eta$ vs. $\phi$ indicates that the tracks are reconstructed in the expected region
of the ATLAS detector. The most probable number of tracks per event is 1 and events with as many as 10 tracks are collected.
The $p_\mathrm{T}$ distribution is peaked just above the FTK $p_\mathrm{T}$ threshold of 1 GeV and falls off rapidly at high $p_\mathrm{T}$, as expected.
Figure~\ref{fig:purity} shows the fraction of FTK tracks matched to selected offline tracks vs. the FTK track
transverse impact parameter $d_0$. More than 98\% of tracks with small $d_0$ are matched to offline tracks,
decreasing to 95\% for tracks with $d_0 =0.5$ mm. This indicates that the majority of tracks reconstructed by
FTK are good tracks corresponding to real particles, as opposed to fake tracks from random combinatoric assignments of hits.
Figure~\ref{fig:resolutions} shows the track parameter resolutions for $p_\mathrm{T}$, $\eta$, and $\phi$.
These results demonstrate that FTK is reconstructing tracks with the correct momentum and directions.
The resolutions are well-modelled using full functional emulation of FTK. The FTK refit, which uses a
full helix fit rather than a linear approximation and also includes additional corrections to the hit position
including that for Lorentz Angle, improves the track parameter resolutions by 10--20\%.

\section{Coping with changing conditions}
\label{sec:conditions}

One of the challenges for operating FTK stably during Run 3 data-taking is the ability to cope with changing conditions.
For example, the beamspot $x$ and $y$ positions may shift over the course of a run, typically by less than 100 $\mu$m.
These changing conditions may affect the sectors and constants and patterns and degrade the tracking performance.

\begin{figure}[!t]
  \centering
  \subfloat[]{\includegraphics[width=0.45\linewidth]{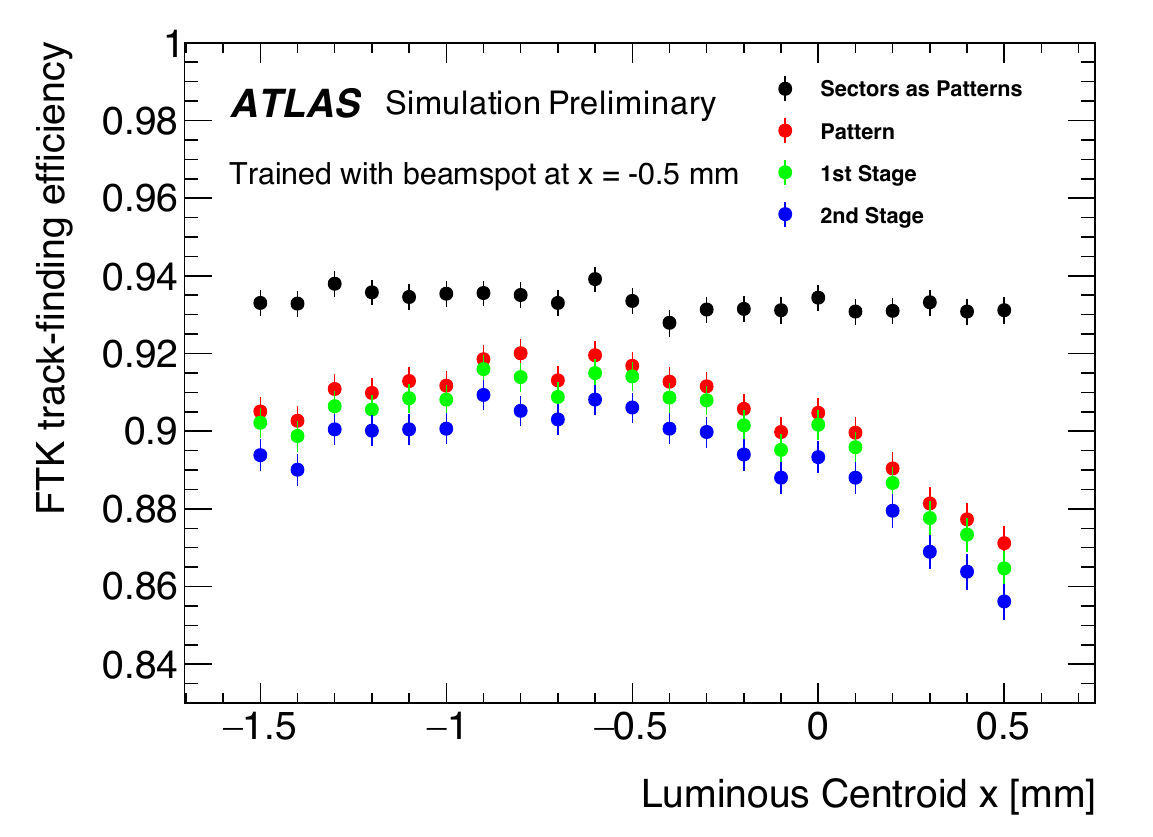}}
  \qquad
  \subfloat[]{\includegraphics[width=0.45\linewidth]{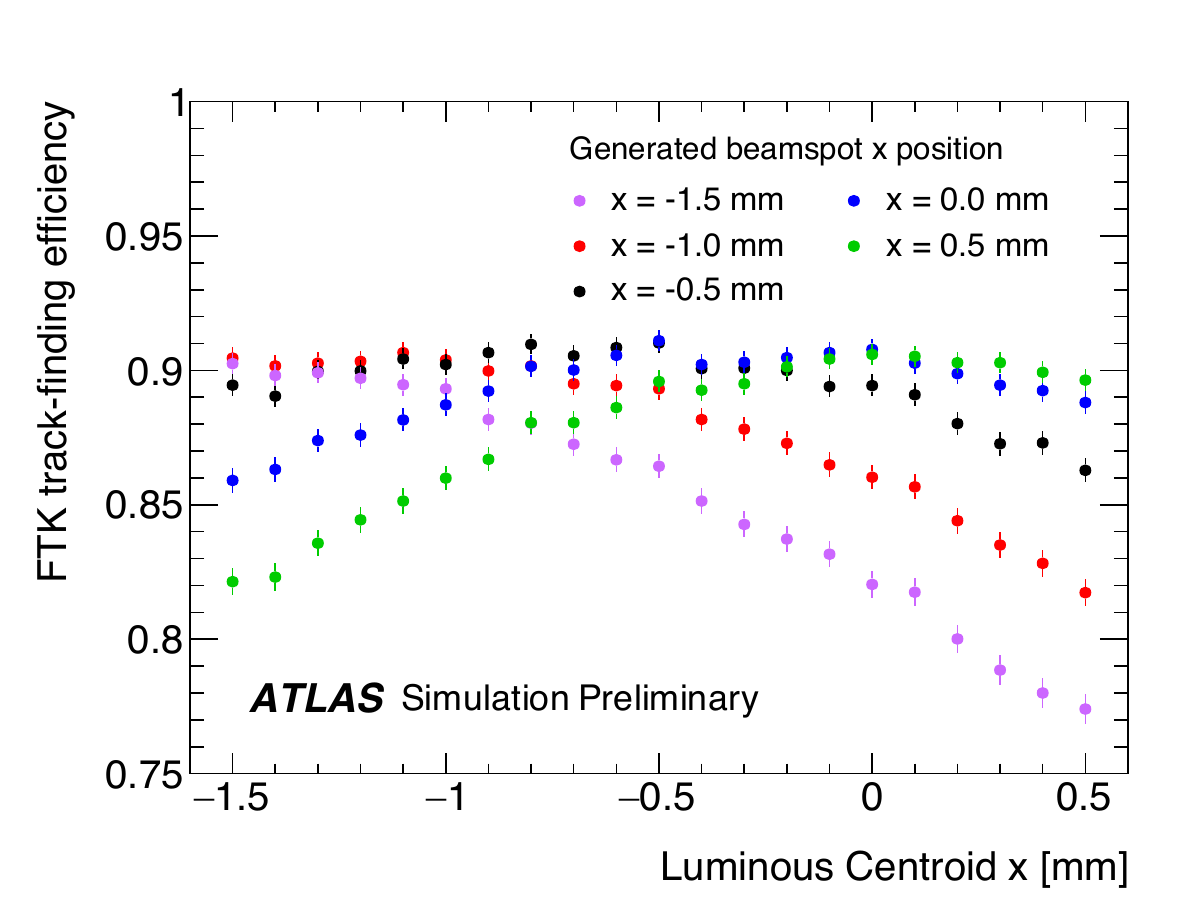}}
  \caption{The FTK track-finding efficiencies vs. beamspot $x$ position~\cite{cite:twiki}.
    The expected FTK tracking efficiency with respect to MC truth from functional emulation of the full FTK system using 
    FTKSim, vs. the x position of the luminous centroid of the beamspot. A fully simulated sample of muons with longitudinal 
    impact parameter $|z_0| < 110$ cm and pseudo-rapidity $|\eta| < 2.5$ is used. 
    Left: the efficiency when treating FTK sectors 
    as patterns (black points) is compared to the efficiency for the nominal patterns (red points) and the efficiencies after 
    the first (green points) and second (blue points) tracking stages. The nominal beamspot used for training the sectors 
    and constants and patterns has $x = -0.5$ mm and $y = -0.9$ mm. 
    Right: the colored dots represent patterns trained with the quoted beamspot x position. 
  }
  \label{fig:conditions}
\end{figure}

To determine the optimal strategy for adapting to changing conditions, the FTK tracking efficiency after several
sequential track-finding steps is evaluated as a function of the beamspot $x$ position, as shown in Figure~\ref{fig:conditions}(a).
Because the efficiency for a track to fall within a defined sector (Sectors as Patterns) does not depend on the beamspot
position, only one set of sectors and constants is needed to maintain high tracking efficiency.
The efficiency for a track to fall within a defined pattern decreases by up to 2\% for beamspot $x$ position shifts
of around 0.5 mm. Therefore, a set of patterns can be produced with beamspot $x$ and $y$ positions shifted in steps of 0.5 mm
to maintain high tracking efficiency, as shown in Figure~\ref{fig:conditions}(b). In Run 3, we therefore plan to use one
set of sectors and constants and produce a set of patterns with generated beamspots every 0.5 mm in $x$ and $y$ position
that can be used in case the beamspot position shifts.

\section{Triggering on long-lived particles}
\label{sec:llp}

\begin{figure}[!t]
  \centering
  \subfloat[]{\includegraphics[width=0.45\linewidth]{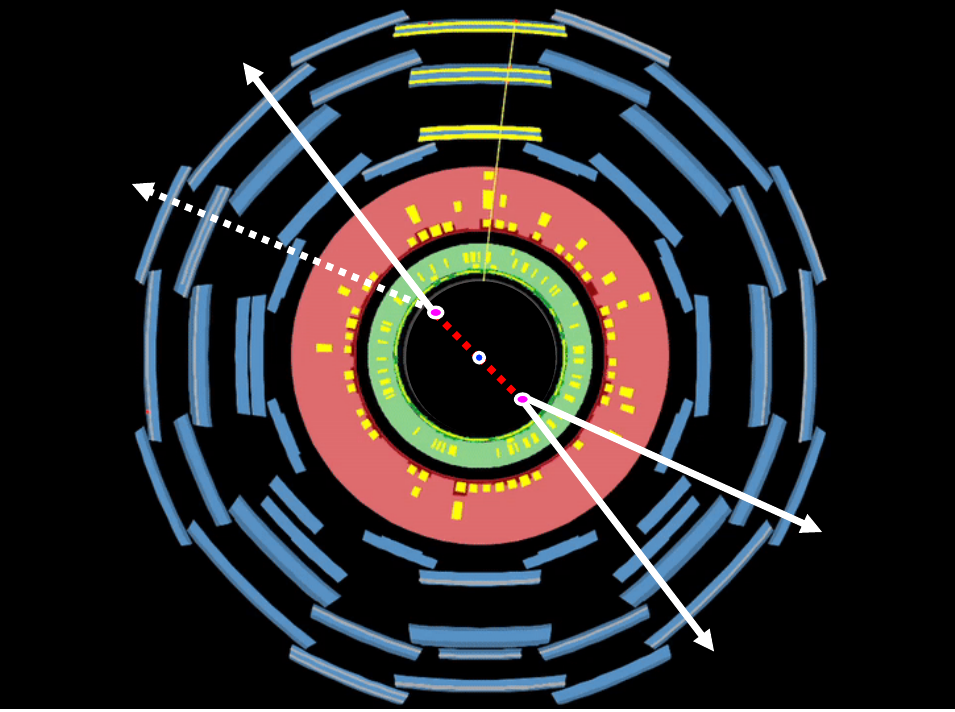}}
  \qquad
  \subfloat[]{\includegraphics[width=0.45\linewidth]{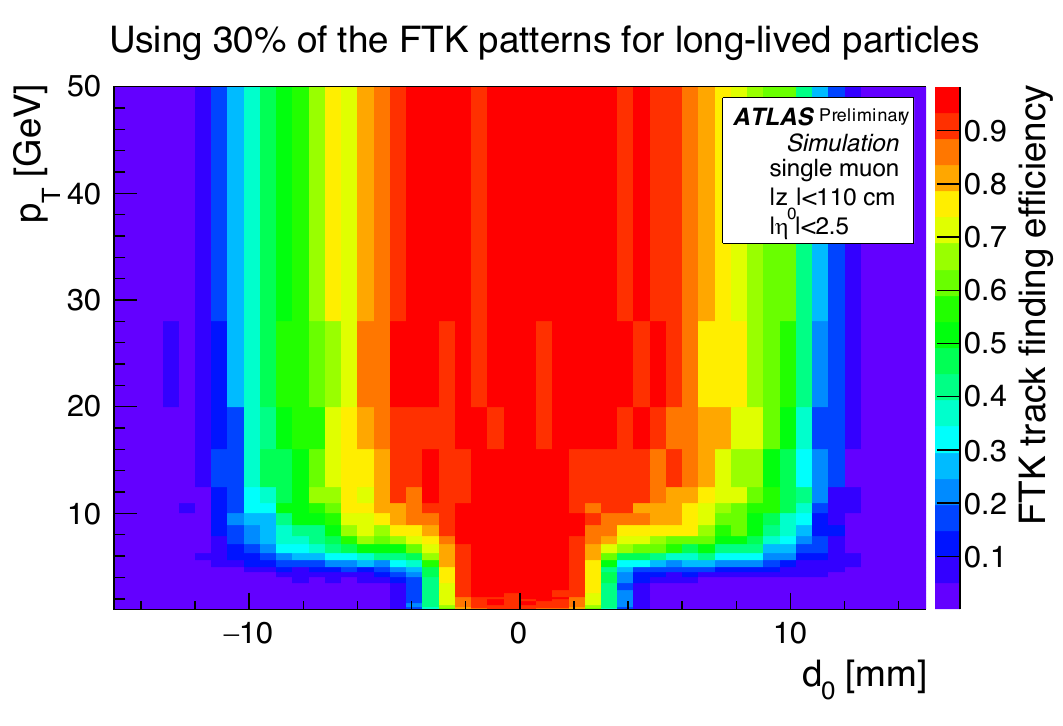}}
  \caption{(a) Event display for a hypothetical signal event with two long-lived particles decaying to 1 or 2 displaced tracks.
(b) The FTK track-finding efficiency vs. track $p_\mathrm{T}$ and transverse impact parameter $d_0$, for a specialized pattern bank
with 30\% of the patterns dedicated to energetic displaced tracks~\cite{cite:twiki}.}
  \label{fig:llp}
\end{figure}

Long-lived particles (LLPs) are theoretically well-motivated in a wide variety of beyond-the-standard model scenarios,
including supersymmetry, hidden valley models, and scenarios with extra dimensions of space-time, but are not yet exhaustively
explored. Scenarios with LLPs represent an excellent physics target for Run 3, in which the collision energy and luminosity
will not increase by large factors, and novel approaches are thus needed to significantly extend the sensitivity beyond Run 2. 
The FTK full-event tracking allows for novel strategies that trigger directly on the displaced tracks that are characteristic
signatures in many scenarios with LLPs, as shown in Figure~\ref{fig:llp}(a). The standard FTK patterns cover tracks with $d_0<2$ mm,
limiting the sensitivity for LLPs that travel several millimeters before decaying. Extending the coverage to larger $d_0$ values
for all tracks would quickly increase the number of patterns beyond the hardware constraints. However, an acceptable increase
in the number of patterns can be obtained by only extending the $d_0$ coverage for high-momentum tracks.
The efficiency for a specialized pattern bank with 30\% of the patterns dedicated to high-momentum high $d_0$ tracks is
presented in Figure~\ref{fig:llp}(b). This bank extends the coverage to tracks with $2<d_0<10$ mm 
without significantly degrading the prompt tracking efficiency.

%Replace the text \cite{example}, Figure~\ref{fig:picture},
%Figures~\ref{fig:pictures}(a--b) and Table~\ref{tab:table1}.
 
%%%%%%%%%%%%%%%%%%%%%%%%%%%%%%%%%%%%%%%%%%%%%%%%%%%%%%%%%%%%%%%%%%%%%%%%%
%%   use this format to include an .eps or .pdf figure(s) into your paper
%%%%%%%%%%%%%%%%%%%%%%%%%%%%%%%%%%%%%%%%%%%%%%%%%%%%%%%%%%%%%%%%%%%%%%%%%

% set the height or width as you prefer

%%%%%%%%%%%%%%%%%%%%%%%%%%%%%%%%%%%%%%%%%%%%%%%%%%%%%%%%%%%%%%%%%%%%%%%%%
%% use this format to include a LaTeX table into your proceedings
%%%%%%%%%%%%%%%%%%%%%%%%%%%%%%%%%%%%%%%%%%%%%%%%%%%%%%%%%%%%%%%%%%%%%%%%%

%%%%%%%%%%%%%%%%%%%%%%%%%%%%%%%%%%%%%%%%%%%%%%%%%%%%%%%%%%%%%%%%%%%%%%%%%%%

\section{Conclusions}
\label{sec:conclusions}

The Fast TracKer is currently being installed and commissioned in ATLAS. First tracking
performance results using a full 12-layer FTK Slice covering an $\eta-\phi$ region of the ATLAS detector
have been presented, based on a sample of approximately 0.5M tracks collected in October 2018.
FTK is capable of producing tracks with the correct momentum and direction.
The resolutions on the transverse momentum and $\eta$ and $\phi$ angles match expectations from full functional emulation using FTKSim,
and 98\% of the FTK tracks are matched to tracks from the default offline reconstruction.
In parallel, groundwork is being laid for
stable FTK running in Run 3. A fast simulation has been developed that uses efficiency
weights and track parameter smearings to significantly reduce computing time with respect to FTKSim.
The strategy for coping with changing beamspot position has been determined, by generating
patterns with beamspot positions spaced 0.5~mm apart in the $x$-$y$ plane.
A special pattern bank for triggering on displaced tracks from long-lived particles has been developed
that extends the sensitivity to high momentum tracks with $2<d_0<10$~mm, without degrading the prompt tracking efficiency.

%%%%%%%%%%%%%%%%%%%%%%%%%%%%%%%%%%%%%%%%%%%%%%%%%%%%%%%%%%%%%%%%%%%%%%%%%%%  

%%  if necessary
%\Acknowledgements
%I am grateful to XYZ for fruitful discussions. Replace the text.

%%%%%%%%%%%%%%%%%%%%%%%%%%%%%%%%%%%%%%%%%%%%%%%%%%%%%%%%%%%%%%%%%%%%%%%%%%%

%%%%%%%%%%%%%%%%%%%%%%%%%%%%%%%%%%%%%%%%%%%%%%%%%%%%%%%%%%%%%%%%%%%%%%%%%%%

\end{document}